# Direct visualization of charge transport in suspended (or free-standing) DNA strands by low-energy electron microscopy

Tatiana Latychevskaia[1], Conrad Escher[1], William Andregg[2], Michael Andregg[2] and Hans-Werner Fink[1]

[1]Physics Department, University of Zurich, Winterthurerstrasse 190, 8057 Zurich, Switzerland

[2]Halcyon Molecular, 505 Penobscot Drive, Redwood City, CA 94063, USA

*Corresponding author: tatiana@physik.uzh.ch

**ABSTRACT**

Low-energy electrons offer a unique possibility for long exposure imaging of individual biomolecules without significant radiation damage. In addition, low-energy electrons exhibit high sensitivity to local potentials and thus can be employed for imaging charges as small as a fraction of one elementary charge. The combination of these properties makes low-energy electrons an exciting tool for imaging charge transport in individual biomolecules. Here we demonstrate the imaging of individual deoxyribonucleic acid (DNA) molecules at the resolution of about 1 nm with simultaneous imaging of the charging of the DNA molecules that is of the order of less than one elementary charge per nanometer. The cross-correlation analysis performed on different sections of the DNA network reveals that the charge redistribution between the two regions is correlated. Thus, low-energy electron microscopy is capable to provide simultaneous imaging of macromolecular structure and its charge distribution which can be beneficial for imaging and constructing nano-bio-sensors.

**MAIN TEXT**

**INTRODUCTION**

Charge transport through deoxyribonucleic acid (DNA) molecules has been a highly interesting (but also controversial) subject over the past few decades in view of the potential for building bio-nano-electronic devices[1-3]. The reports about the electronic properties of DNA are highly controversial[4]. Time-resolved experiments reported ultrafast electron transfer in double stranded DNA (dsDNA) with time constants of 5 ps and 75 ps over 10 – 17 Å distance[5]. Kasumov et al reported that dsDNA molecule exhibited ohmic behaviour between room temperature and 1 Kelvin with resistance per molecule less than 100 kiloohm, and below 1 K proximity-induced superconductivity was observed[6]. Okahata et al investigated electrical conductivity of in DNA-lipid complex film, where they measured electrical current in dsDNA- and almost no electrical current in single stranded DNA (ssDNA)-lipid complex films; these results imply that the conduction of ssDNA could be much less than the conduction of dsDNA[7]. Some DNA transport measurements indicated that DNA molecules could be conductive[7-9]. Fink et al measured that a DNA molecule could act as a semiconductor exhibiting a resistivity of about 1 mΩ/cm[8]. Yoo et al reported that poly(dA)-poly(dT) behaved as an n-type semiconductor, whereas poly(dG)-poly(dC) behaved as a p-type semiconductor[9]. Other experiments indicated that DNA could be insulating[10-13]. An overview of possible mechanisms of charge transport through DNA is provided by Generaux and Barton[1].

Low-energy electrons with kinetic energies in the range 30 - 250 eV provide a unique type of radiation which causes no significant radiation damage to biological molecules, as was exemplified by continuous exposure of individual DNA molecules to low-energy electrons for 70 min, without noticeable change in their interference pattern (hologram)[14]. The number of electrons required to acquire a single 20 ms low-energy electron hologram at 1 nm resolution amounts to about 250 electrons per 1 Å$^2$, which translates into a radiation dose of $4.58 \times 10^{11}$ Gray. This radiation dose exceeds the maximum tolerable dose for high-energy electrons and X-ray imaging by about a factor of $10^4$. The details of this calculations are provided in the Supplementary Information.

During the past two decades, low-energy electrons have successfully been applied for imaging of individual biological molecules, including: purple protein membrane[15], DNA molecules[16-18], phthalocyaninato polysiloxane molecules[19], the tobacco mosaic virus[20], a bacteriophage[21], ferritin[22] and individual proteins (bovine serum albumin, cytochrome C and hemoglobin)[23]. Most of these results were obtained by imaging individual molecules stretched over holes in carbon films[15-21]. However, such a sample arrangement creates an unwanted so-called biprism effect. Such biprism effect occurs when the electron wave passes by a positively charged wire, so that the electrons are deflected towards the wire[24]. In light optics such effect can be created by adding a biprism phase distribution into the wavefront. In low-energy electron imaging even if the fiber is not charged, such biprism effect can occur due to the bending of the potential around the molecule resulting in a deflection of the electron trajectories similar to as if the molecule was charged[20]. The biprism effect complicates the interpretation of the data record[25]. Biprism effects can be reduced if the individual molecules are stretched over smaller holes in a carbon film. Simulations performed by Weierstall et al[20] demonstrated that stretching an 18 nm fiber over a 100 nm instead of typical 1-2 micron holes, successfully suppresses the biprism effect. In this study we demonstrate low-energy electron imaging of individual DNA molecules that are stretched over holes in lacey carbon with hole sizes of just tens of nanometers.

Low-energy electrons exhibit high sensitivity to local potentials[26] allowing imaging individual charges as small as a fraction of an elementary charge[27-29]. This is why low-energy electron imaging is a unique tool to probe charge effects in DNA molecules at high spatial resolution and at high sensitivity to the smallest amount of charge.

## SAMPLE PREPARATION

The sample consisted of individual single-stranded DNA (ssDNA) strands as well as bundles thereof stretched over holes in lacey carbon. The thymine bases in ssDNA were labelled with osmium atoms by staining ssDNA with a thymidine-selective osmium tetroxide 2-29 bipyridine (osbipy) contrast-enhancing label. The ssDNA strands were prepared by the ''molecular threading'' method - a surface independent tip-based method for stretching and depositing single and double-stranded DNA molecules[30]. DNA was stretched into air at a liquid-air interface and subsequently deposited onto a dry substrate isolated from solution. A fluorescence microscopy image of such sample is shown in Fig. 1(a).

## EXPERIMENTAL SETUP

The low-energy electron microscope employed in this study has been described in details in previous publications [16,18,21-23,31] and is schematically shown in Fig. 1(b). The source of the coherent electron beam was a sharp W(111) tip and the electrons were extracted by field emission[32]. The position of the tip was controlled by a 3-axis piezo-manipulator with nanometer precision. The wave transmitted through the sample propagated to the detector unit where the interference pattern is acquired, formed by superposition of the scattered with the non-scattered (reference) wave, constituting an in-line hologram[33,34]. The detector unit consisted of a microchannel plate (MCP), a phosphor screen, and a digital camera. A typical inline hologram of ssDNA fibers acquired with this setup is shown in Fig. 1(c), exhibiting ssDNA fibers perfectly stretched over the lacey carbon support. Examples of ssDNA holograms at different magnifications are provided in the Supplementary Information.

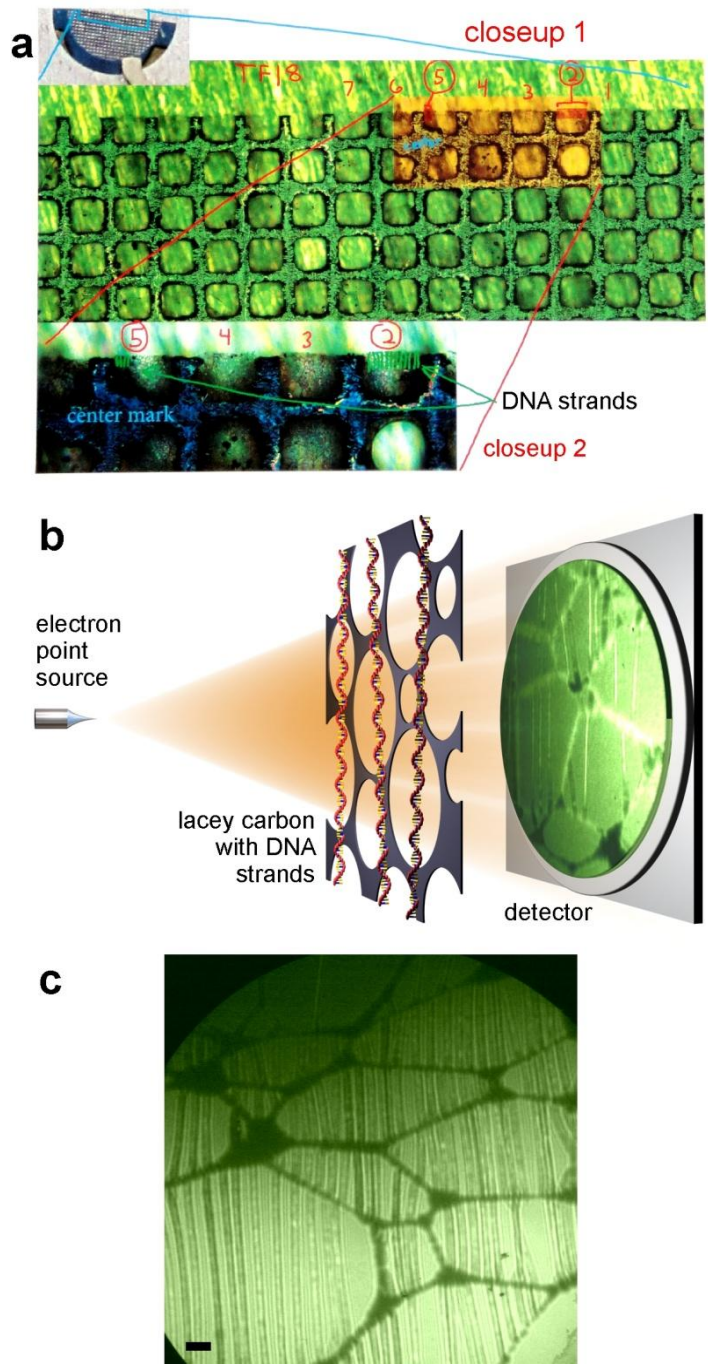

Fig. 1 Imaging single-stranded DNA molecules by low-energy electron microscopy. (a) The sample imaged by fluorescence optical microscopy at different magnifications, where two compartments (2 and 5) containing individual DNA strands are indicated. (b) Schematics of the low-energy electron microscope where the sample can be imaged at different magnification, the electron source-to-sample distance can be varied from tens of nanometers to a few microns. (c) An in-line hologram of ssDNA fibers in compartment 2 as labelled in (a), acquired with electrons of 188 eV kinetic energy, at the distance between the electron source and the sample of about 100 μm. The scalebar corresponds to 100 nm.

## RESULTS

### Visualisation of charge redistribution

Figure 2 shows low-energy electron microscopy images of ssDNA. In each experiment, a sequence of holograms was recorded at a standard video rate (25 frames per second). An individual hologram (frame) is shown in Figure 2(a), where one can see ssDNA fibers stretched over holes in lacey carbon. Some ssDNA fibers exhibit brighter or darker regions, which is an indication of negative (darker regions) and positive (brighter regions) charge. The presence of charge affects the electron trajectories: negative charge deflects passing electrons away from the charge, while positive charge deflects passing electrons towards the charge thus creating a lens-like effect. A sequence of holograms of the same sample as shown in Fig. 2 is provided as a movie in Supplementary Movie 1. Part of the imaged sample contained only the carbon fibers without DNA fibers, marked by the yellow rectangles in Fig. 2(a) and (b). From the Supplementary Movie 1, it is apparent that the intensity contrast is only varying along the DNA fibers, and no intensity contrast variation was observed along the carbon fibers in the regions marked by the yellow rectangle. We therefore can assume that the charge redistribution occurs only along the DNA fibers.

A somewhat blurry appearance of the experimental images (in Fig. 2, 3, S2 and Supplementary Movies 1 and 2) can be explained as follows. When the electron source or the sample is laterally shifting, the image on the detector (hologram) is laterally shifting by the same amount of shift multiplied with the magnification of the system. For example, a sample shift by 1 Ångstrom leads to a hologram shift by 10 micron at a typical magnification of $10^5$. In addition, the fibers in the lacey samples are not mechanically rigid and can also exhibit deflection in the axial direction, which affects the magnification of the resulting images. Since the lateral and axial positions of the fibers are continuously varying (as it can be observed in Supplementary Movies 1 and 2), the resulting hologram of the fiber, averaged over the time period corresponding to the single frame acquisition time, appears blurry.

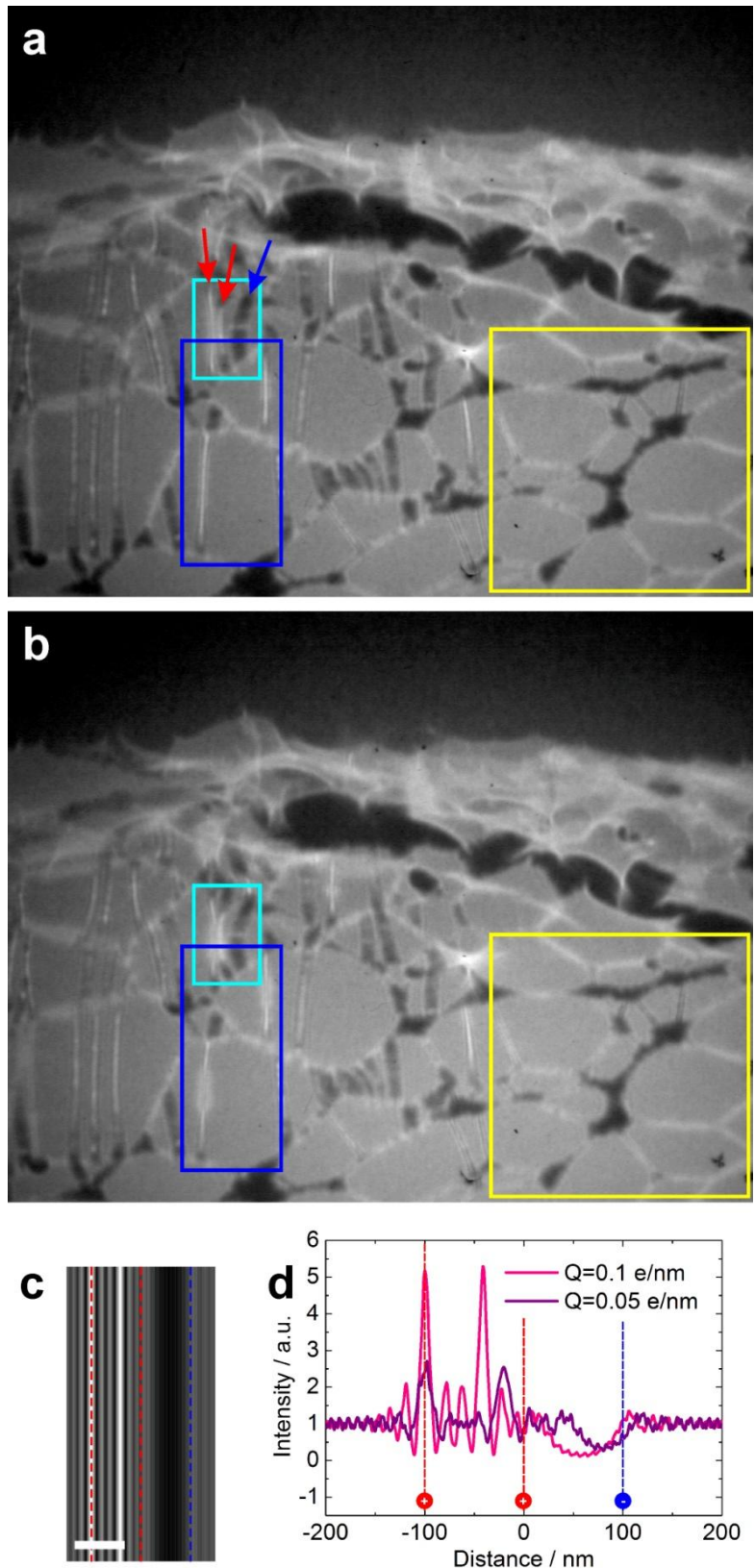

Fig. 2. Low-energy electron microscopy of ssDNA. (a) Single hologram of ssDNA fibers suspended over holes in lacey carbon acquired with electrons of 215 eV energy and a 10 nA current. (b) Hologram of the same region as in (a), acquired 600 ms later. A complete movie, showing the time-dependent behaviour of this region under continuous

exposure to electrons is provided as the Supplementary Movie 1. (c) Simulated hologram of three 2 nm thick fibers with the linear charge of the fibers is $|Q|=0.1$ e/nm, $+|Q|$ for the left and the center fibers and $-|Q|$ for the right fiber. The distance between the fibers is 100 nm, the electron energy is 215 eV and the distance between the source and the sample is 10 μm. The scalebar corresponds to 100 nm. (e) Intensity profiles through the simulated holograms of three charged fibers with $|Q|=0.1$ e/nm and $|Q|=0.05$ e/nm. The actual positions of the charged fibers are indicated by the dashed lines in (c) and (d).

**Quantitative estimation of charges**

The high sensitivity of low-energy electrons to local potentials allows detecting smallest charges[28,29]. The effect of electron trajectories deflections due to charges and their redistribution in biological samples is often discussed in biological single particle transmission electron microscopy imaging[35,36]. In low-energy electron holography, due to this effect, the resulting in-line holograms of the molecules are affected by distortions, as shown in Fig. 2(a) – (b) in the cyan rectangle. The three ssDNA fibers in Fig. 2(a) – (b) in the cyan rectangle, indicated with two red and one blue arrows, are not physically bent towards each other, this is a distortion in the in-line hologram image created by the electron trajectories that have been deflected due to the charges present on the molecules, as indicated by the bright (due to a positive charge) and dark (due to a negative charge) appearance of the ssDNA fibers. To illustrate this "bending" effect, a similar situation of three charged fibers separated by 100 nm was simulated. A linear charge distributed over the fiber was assumed, which creates a biprism phase-shifting distribution; the procedure of hologram simulations of fibers with linear charge is explained in ref[25]. A simulated hologram when the linear charge of the fibers is $Q=0.1$ e/nm, $+|Q|$ for the left and the center fibers and $-|Q|$ for the right fiber, is shown in Fig 2(c). The intensity profile through the simulated holograms at $Q=0.05$ e/nm and $Q=0.1$ e/nm are shown in Fig 2(d), together with the indicated actual positions of the fibers. It is apparent that the maxima and minima of the intensity in the holograms are significantly shifted, almost by 50% from the original fiber positions. It is also remarkable that such a huge shift is caused by such small charges, illustrating the high sensitivity of low-energy electrons to local charges.

**Correlated charge redistribution**

Figure 3 shows a study of the time evolution of the intensity fluctuations in the hologram of an ssDNA. A region selected for the analysis is indicated in Fig. 2(a) by the blue rectangle. The bottom left ssDNA fiber exhibits a biprism effect which is an indication of positive charging. The first 10 holograms (frames) of the region are shown in Fig. 3(a). The normalized intensity as a function of time (frame number) at two selected sub-regions is shown in Fig. 3(b). The intensity as a function of time (frame number) at two adjacent sub-regions is shown in Fig. 3(c). The normalized intensity values at the selected sub-regions (indicated by the blue and the lilac arrows, respectively) were calculated as follows. The intensity is averaged over a 27 × 27 nm$^2$ area in the sub-region indicated by the solid arrow, giving $I_i$, $i$ =1, 2, respectively. An averaged intensity over a 27 × 27 nm$^2$ area in the reference sub-region indicated by the dotted arrow was also calculated, giving $I_i^{(\text{ref})}$, $i$ =1, 2, respectively. The normalized intensity was calculated as $I_i / I_i^{(\text{ref})}$ $i$ =1, 2, respectively. The cross-correlation function (CCF) between the normalized intensities at the sub-regions at 1 and 2 is shown in Fig. 3(d). The CCF$_{\text{ref}}$ of the intensities at the reference sub-regions is shown in Fig. 3(e), exhibiting a periodical fluctuation and a broad maximum caused by the intensity drop between frames 45 and 115 (as shown in Fig.3(c)). These CCF$_{\text{ref}}$ features are not present in CCF of the normalized intensities shown in Fig. 3(d), though the CCF of the normalized intensities still shows some oscillations. Another remarkable difference between the two CCFs is that the CCF of the normalized intensities exhibits a minimum at frame=0, thus indicating that there is a time shift between the two intensity distributions, as if the two intensity distributions can be described by cosine functions that are shifted by π relatively to one another. The CCF distribution and the time shift implies that the charge redistribution between the two regions is not completely random but correlated.

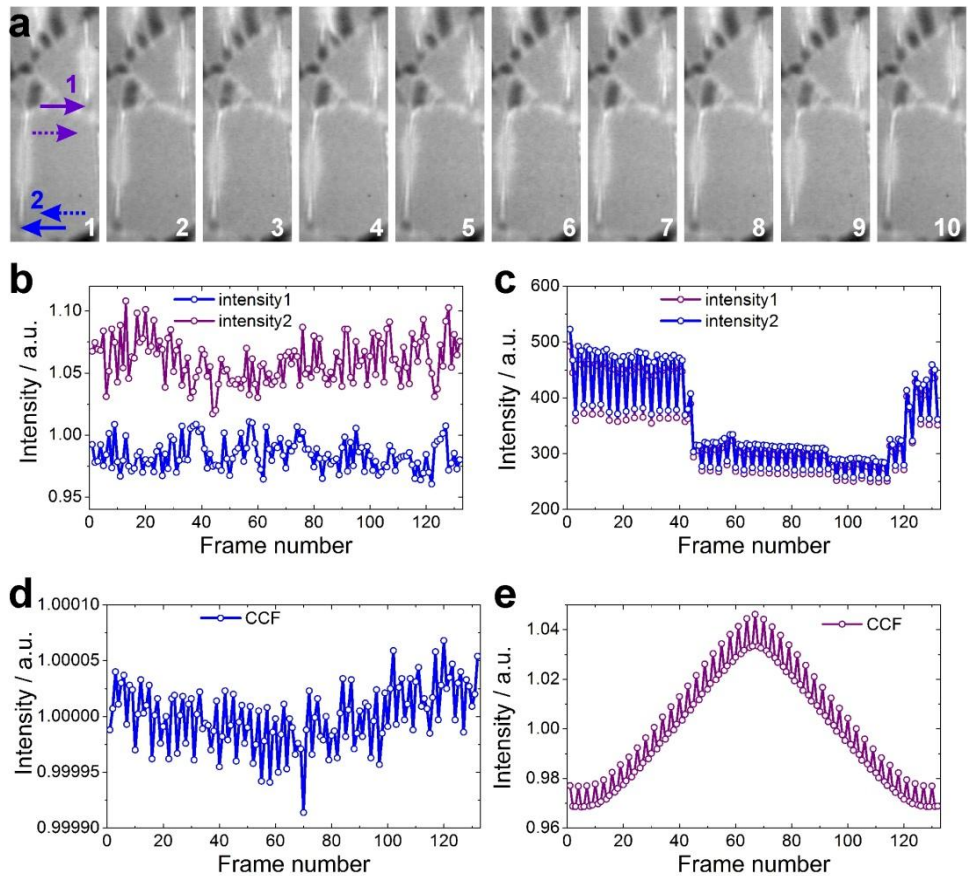

Fig. 3. Time evolution of intensity fluctuations in a hologram of ssDNAs. (a) Selected area in the first ten frames of a sequence of 132 frames. The selected area is shown in Fig.2(a) in the blue rectangle. The frame number is shown in the bottom right corner of each frame. Two selected sub-regions are indicated by the blue and the lilac arrows, respectively. Two corresponding reference regions are indicated by the blue and the lilac dotted arrows, respectively. (b) Normalized intensity as a function of time (frame number) at the sub-regions 1 and 2 as indicated by the colored arrows in (a). (c) Intensity as a function of time (frame number) at the references sub-regions 1 and 2 as indicated by the dotted colored arrows in (a). (d) Cross-correlation function calculated between intensities at the sub-regions 1 and 2 (CCF). (e) Cross-correlation function of the intensities at the reference sub-regions (CCF$_{ref}$).

**DNA molecules structure reconstruction**

Figure 4 shows in-line low-energy electron holograms of ssDNA fibers and their reconstructions. In the holograms, one can notice bright blobs along the fibers (in particular in Fig. 4b), which can be associated with a small amount of localized positive charges. These charges demonstrate small oscillation-like movements around their position over time, as can be viewed in the Supplementary Movie 2. The reconstructions were obtained by numerical procedure as described elsewhere[37]. The width of the fibers were evaluated from the reconstructions and amount for fiber D to: 4.57 ± 0.51 nm, and fiber E to: 6.60 ± 0.51 nm. This implies that fibers D and E are rather a bundle than individual ssDNA molecules. The individual thymine bases that are labelled by Osmium cannot be resolved in the obtained hologram reconstructions.

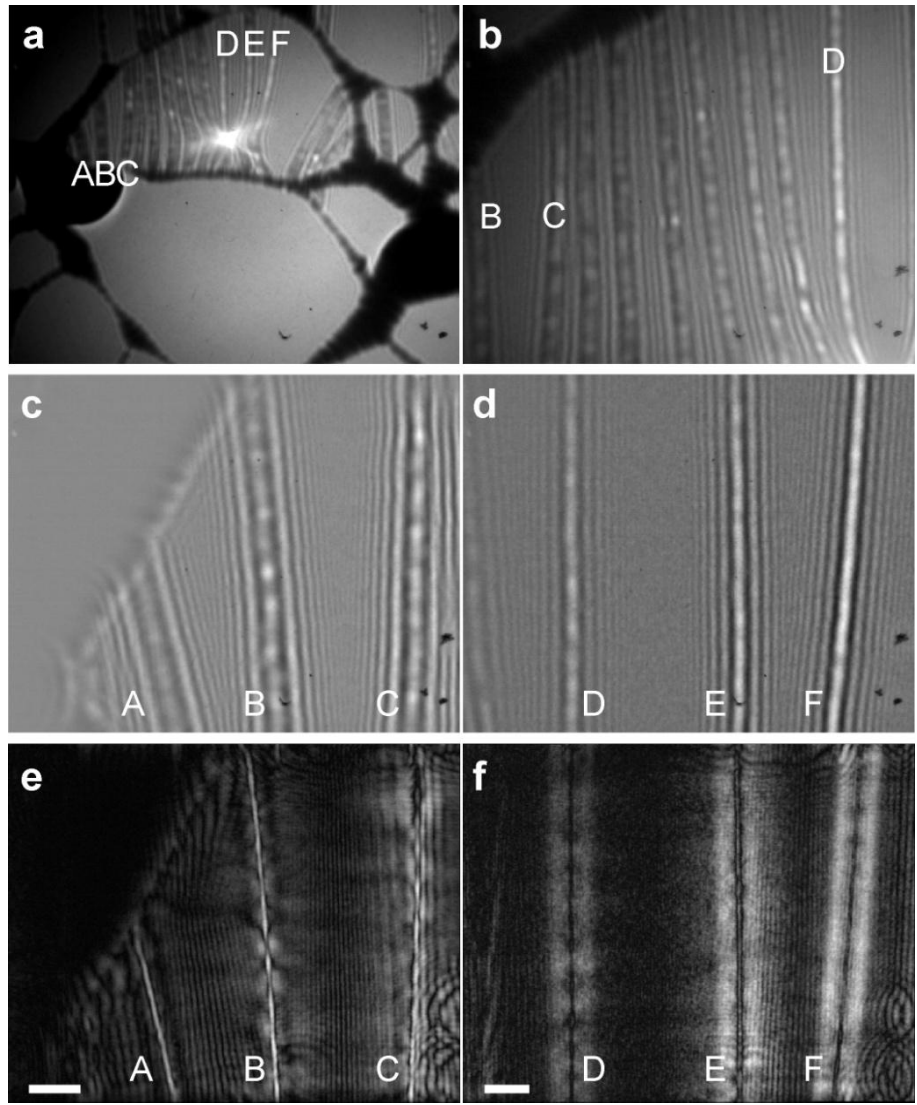

Fig. 4. Low-energy electron holograms of ssDNA fibers and their reconstructions. (a) Overview image of a section of ssDNA fibers suspended over holes in lacey carbon, acquired with electrons of 154 eV energy and a 5 nA current. (b) Magnified image of the same region, acquired at a shorter electron source- to-sample distance, with electrons of 110 eV energy and a 400pA current. Time-dependent behaviour of this region under continuous exposure to electrons is provided as the Supplementary Movie 2. (c) and (d) Magnified regions with ssDNA fibers A, B, C, D, E, and F acquired with electrons of 110 eV energy and a 350 pA current, at a distance between the source and the sample of 4.9 μm and 5.6 μm, respectively. (e) and (f) sample distributions obtained by reconstruction of the holograms shown in (c) and (d). The scalebars in (e) and (f) correspond to 50 nm.

Figure 4a shows an in-line low-energy electron hologram of ssDNA fibers with three fibers seemingly merging together into a bright spot. This is another manifestation of the "bending effect" discussed above. Due to the presence of a strong positive charge on the middle fiber D, the electron trajectories are bent towards the positive charge, thus creating on the detector an image of bent neighbouring fibers.

**Discussion and Conclusions**

In conclusion, our results demonstrate that low-energy electron microscopy allows imaging of structure of biological macromolecules at about 1 nm spatial resolution and simultaneous imaging of its charge distribution at the resolution of sub elementary charges. Although biological samples are known to undergo charging effects under electron imaging, only with low-energy electrons this charging can directly be visualized.

Our experiments demonstrated that a region of the sample which contains only carbon fibers does not exhibit such intensity fluctuations as the nearby region of the sample which contains DNA molecules. We therefore conclude that we observe charge redistribution within DNA. The subject of DNA conductivity is highly controversial and we do not have an explanation about the exact mechanism of charge redistribution in DNA. We can only speculate that in our experiments the situation could be similar to that in graphene: although graphene is highly conductive in theory, in practise graphene has defects (missing atoms, adatoms etc) which affect the conductivity severely. Adatoms on graphene can produce a local charge transfer[38], where the local charge can exhibit fluctuations in time[28].

We show that a charged ssDNA fiber can lead to a shift by a few tens on nanometers of the fiber's image (hologram) from its original position. For the ssDNA network, the cross-correlation analysis reveals that the charge redistribution between the two regions in the network that are tens of nanometers apart is not completely random but correlated. This result can potentially be useful for employing ssDNA networks in molecular electronics. Overall, low-energy electron microscopy offers a unique visualization tool for studying the charge transport in DNA and other biomolecules which could be potentially applied for the construction of nano-bio-sensors.

**Author Contributions**

W. A. and M. A. prepared the ssDNA samples, H. W. F and C. E. performed the experiments, T.L. analyzed the data and wrote the initial draft of the paper, all authors participated in discussing and finalizing the manuscript.

# Supplementary Information for

# Direct visualization of charge transport in suspended (or free-standing) DNA strands by low-energy electron microscopy

Tatiana Latychevskaia[1], Conrad Escher[1], William Andregg[2], Michael Andregg[2] and Hans-Werner Fink[1]

[1]Physics Department, University of Zurich, Winterthurerstrasse 190, 8057 Zurich, Switzerland

[2]Halcyon Molecular, 505 Penobscot Drive, Redwood City, CA 94063, USA

**Electron dose and radiation damage**

For an electric current of 200 nA, there are $1.248 \times 10^{12}$ electrons flowing per second. Assuming that all these electrons illuminate an area of $1 \times 1$ um$^2$, we obtain $1.248 \times 10^4$ electrons per second per 1 Å$^2$. This gives 250 electrons per 1 Å$^2$ for a 20 ms single hologram frame. For a typical electron dose of $1 \times 10^4$ per 1 Å$^2$ ($1 \times 10^6$ per 1 nm$^2$) deposited by 100 eV electrons per second, the radiation dose amounts to[1]:

$$R = \frac{(\text{dose in } e/\text{nm}^2) \times (\text{energy in eV})}{(\text{penetration depth in mm}) \times (\text{density in g/cm}^3)} \cdot 0.16 = \frac{1 \cdot 10^6 \cdot 100}{5 \cdot 10^{-7} \cdot 1.4} \cdot 0.16 = 2.29 \cdot 10^{13} \text{ Gray},$$

where we assumed the penetration depth of 5 Å and a typical density of biological specimen of 1.4 g/cm$^3$. This thus amounts to $4.58 \times 10^{11}$ Gray per one hologram. Taking into account that at the modest, the resolution of about 1 nm can be achieved when imaging with 100 eV low-energy electrons, the radiation dose exceeds the maximum tolerable dose for high-energy electrons and X-ray by about a factor of $10^4$, as illustrated in Fig. S1. It has been demonstrated that DNA molecules do not exhibit significant radiation damage, at the resolution of 1 nm, when continuously exposed to low-energy electrons for 70 min[2], thus resulting in a total dose of $9.62 \times 10^{16}$ Gray.

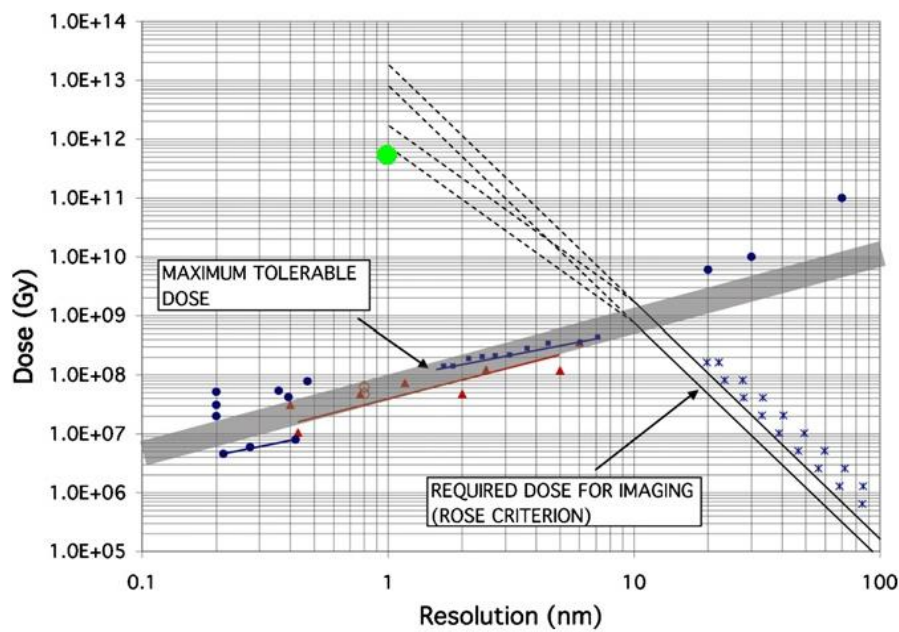

Fig. S1. Graph summarizing information on the required dose for imaging and the maximum tolerable dose. Reprinted from Journal of Electron Spectroscopy and Related Phenomena, Volume 170, Issues 1-3, M.R. Howells, T. Beetz, H.N. Chapman, C. Cui, J.M. Holton, C.J. Jacobsen, J. Kirz, E. Lima, S. Marchesini, H. Miao D. Sayre, D.A. Shapiro, J.C.H. Spence, D. Starodub, "An assessment of the resolution limitation due to radiation-damage in X-ray diffraction microscopy", Pages 4-12, Copyright (2019), with permission from Elsevier. The green dot indicates the radiation dose for low-energy electrons when imaging biological specimen for 20 ms (one hologram) at 1 nm resolution. The rest of the graph is described in as follows[3]. The types of data from the literature are identified by the symbols as follows: filled circles: X-ray crystallography; filled triangles: electron crystallography; open circles: single-particle reconstruction; open triangles: electron tomography; diamonds: soft X-ray microscopy[3]. The required dose for imaging is calculated for a protein of the empirical formula $H_{50}C_{30}N_9O_{10}S_1$ and a density of 1.35 g/cm$^3$ against a background of water for X-ray energies of 1 keV (lower continuous line) and 10 keV (upper continuous line)[3]. The dashed continuations of these lines refer to the transition region from a coherent to an incoherent behaviour. Measurements of the required dose for X-ray imaging are plotted as crosses[3]. The maximum tolerable dose is obtained from a variety of experiments by Howells et al[3].

**Low-energy electron imaging at different magnification**

Figure S2 shows three images of the ssDNA sample acquired in the low-energy electron microscope at different magnification. The magnification is changed by varying the distance between the electron source and the sample.

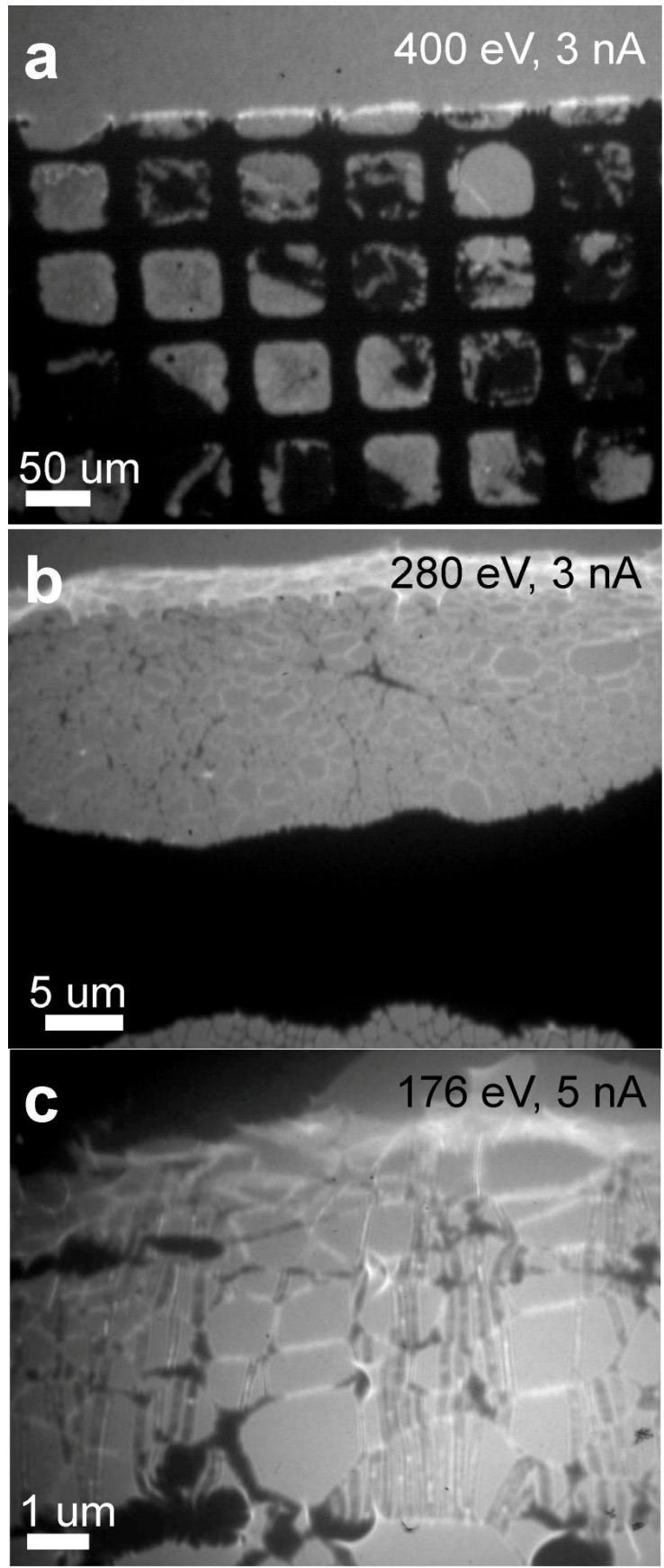

Fig. S2. Images of the ssDNA sample acquired in the low-energy electron microscope at different magnification.